\documentclass[aps,pra,twocolumn,floatfix]{revtex4-2}

\usepackage[T1]{fontenc}
\usepackage{amsfonts,amssymb}
\usepackage{nccmath}
\usepackage{mathtools}
\usepackage{xspace}
\usepackage{tensor}
\usepackage{csquotes}
\usepackage{graphicx}
\usepackage{xcolor}
\usepackage{enumitem}
\usepackage{ctable, multirow, collcell}
\usepackage{soul}
\usepackage{hyperref}
\usepackage[capitalize]{cleveref}
\usepackage{siunitx}

\DeclareSIUnit\Gauss{G}
\DeclareSIUnit{\au}{a.u.\xspace}

\newcommand{\gfull}{\ensuremath{5s^2\,\tensor[^1]{\mathrm{S}}{_0}}\xspace}
\newcommand{\efull}{\ensuremath{5s5p\,\tensor[^3]{\mathrm{P}\!}{_1}}\xspace}
\newcommand{\gshort}{\ensuremath{\tensor[^1]{\mathrm{S}}{_0}}\xspace}
\newcommand{\eshort}{\ensuremath{\tensor[^3]{\mathrm{P}}{_1}}\xspace}

\begin{document}

\title{Differential polarizability at 1064\,nm of the strontium intercombination transition}

\date{\today}

\author{Romaric Journet}
\author{Félix Faisant}
\altaffiliation{RJ and FF have equally contributed to this work.}
\author{Sanghyeop Lee}
\author{Marc Cheneau}
\affiliation{Universit{\'e} Paris Saclay, Institut d'Optique Graduate School, CNRS, Laboratoire Charles Fabry, 91127, Palaiseau, France}

\begin{abstract}
    We measure the scalar, vector and tensor components of the differential dynamic polarizability of the strontium intercombination transition at \qty{1064}{\nm}.
    We compare the experimental values with the theoretical prediction based on the most recently published spectroscopic data, and find a very good agreement.
    We also identify a close-to-circular `magic' polarization where the differential polarizability strictly vanishes, and precisely determine its ellipticity.
    Our work opens new perspectives for laser cooling optically trapped strontium atoms, and provides a new benchmark for atomic models in the near infrared spectral range.
\end{abstract}

\maketitle

\section{Introduction}

Ultracold gases of alkaline-earth and alkaline-earth-like atoms have long been identified as a promising platform for quantum computation and quantum simulation, with distinct advantages compared to alkaline atoms \cite{Gorshkov2010,Daley2011,Mukherjee2011,Banerjee2013,Dunning2016}.
One is the decoupling between the \gshort electronic ground state and the nuclear spin, which opens up the possibility to encode and manipulate well isolated qubits \cite{Barnes2022,Jenkins2022,Huie2023}, or to simulate many-body phenomena with $\text{SU}(n)$ symmetry, where $n$ is the number of nuclear spin states and can be as large as 10 for \nuclide[87]{Sr} \cite{Zhang2014,Goban2018}.
Another is the series of narrow and ultra-narrow transitions from the singlet ground state to the triplet excited states $\tensor[^3]{\mathrm{P}\!}{_J}$, which can be leveraged, for instance, for metrological applications \cite{Ludlow2015,Kluesener2024}, or laser cooling to extremely low temperatures \cite{Stellmer2014}.
Last but not least, they offer the prospect of using the optical transitions of the secondary valence electron to image or confine atoms in a Rydberg state \cite{Pham2022}.

Recently, a new generation of experiments has flourished, which exploit these properties with strontium atoms confined in optical microtraps, whether optical tweezer arrays \cite{Norcia2018,Cooper2018,Jackson2020,Park2022,Urech2022,Hoelzl2023}, optical lattices \cite{Park2022,Buob2024}, or a combination of both \cite{Young2022,Tao2023}.
In order to design such experiments, it is essential to have a good knowledge of the dynamic polarizability at the trapping wavelength.
For instance, the performance of laser cooling on the \gshort--\eshort intercombination transition inside a microtrap crucially depends on the differential dynamic polarizability between the ground and excited states:
When both states have equal polarizabilities, resolved sideband cooling is known to be a powerful technique which can reduce the atomic motion to the vibrational ground state \cite{Norcia2018,Cooper2018};
When the excited state has a larger dynamic polarizability than the ground state, an \emph{attractive} Sisyphus-type cooling mechanism for a red detuning with respect to the free-space transition was predicted decades ago \cite{Taieb1994,Ivanov2011}, and has indeed proven very efficient and applicable to a wide range of trap depths and stiffnesses \cite{Chen2019,Covey2019,Urech2022,Buob2024};
Finally, when the excited state has a smaller polarizability than the ground state, an alternative \emph{repulsive} Sisyphus-type mechanism with blue detuning can also be exploited, under the condition that the atoms never reach the position in space where the laser becomes resonant with the shifted transition \cite{Cooper2018,Hoelzl2023,Tao2023}.

To compute the dynamic polarizability of a given state, one usually sums the contributions of all known transitions connecting to this state with the appropriate transition energies and dipole moments.
These parameters can either be measured, or obtained from ab-initio \cite{Mitroy2010,Safronova2013,Cooper2018} or semi-empirical models \cite{Ruczkowski2016}.
Because the precise determination of the transition dipole moments is a notoriously difficult task, and some discrepancies exist in the literature, it is necessary to cross-check the validity of the computation by looking for singular features such as `tune-out' (zero polarizability) or `magic' (zero differential polarizability) configurations at specific wavelengths \cite{Safronova2015}, and compare the prediction to experimental measurements.
For strontium, magic configurations have been predicted and observed at \qty{515.2}{\nm} \cite{Norcia2018,Cooper2018} and \qty{914}{\nm} \cite{Ido2003} for the intercombination transition, and at \qty{813.4}{\nm} for the \gshort--$\tensor[^3]{\mathrm{P}}{_0}$ clock transition \cite{Takamoto2006,Brusch2006,Boyd2007},
and, most recently, a predicted tune-out configuration has also been observed at \qty{689.2}{\nm} for the ground state \cite{Heinz2020}.

Around \qty{1064}{\nm}, where powerful laser sources are available for trapping atoms, a magic configuration has been predicted for the intercombination transition of strontium \cite[pp.~65-66]{BoydThesis} \cite[pp.~39-40]{MadjarovThesis}, and was supported by early experimental observations \cite[p.~179]{StellmerThesis}.
However, a direct observation was still lacking.
In this work, we have measured the scalar, vector, and tensor components of the differential dynamic polarizability of the strontium intercombination transition at \qty{1064.7}{\nm}.
Our measurements are in excellent agreement with the reference values which we computed using the most recent spectroscopic data available in the literature \cite{NIST-ASD,Ruczkowski2016,Zhou2010,Cooper2018}, which confirms the accuracy of the spectroscopic data.
We have also observed the expected magic polarization for the states with a non-zero magnetic quantum number.
Our results are summarized in \cref{tab:results} at the end of the article.

\section{Theoretical background and prediction for the polarizability%
\label{sec:theory}}

In this section, we quickly introduce the theoretical framework, and present our prediction for the differential dynamic polarizability of the intercombination transition at \qty{1064}{\nm}.
Our presentation closely follows that of \cite{LeKien2013}.
In our context, the electronic eigenstates are uniquely defined by the set of quantum numbers $(\gamma, J, m)$, where $J$ and $m$ correspond, respectively, to the total electronic angular momentum and its projection along the quantization axis, and
$\gamma$ is a shorthand notation for the electronic configuration and spectroscopic term which we will omit in the reminder of the article.
We assume that the quantization axis is provided by a bias magnetic field $\mathbf{B}$.
In the dipole approximation, the interaction between the atomic dipole $\mathbf{d}$ and a laser field with complex electric field amplitude $\mathcal{E}$, polarization $\mathbf{u}$, and angular frequency $\omega$, is described by the operator
\begin{equation}
    \hat{V} = -\frac{1}{2} \mathcal{E} \mathbf{u} \cdot \hat{\mathbf{d}} \, e^{-i\omega t} + \text{h.\,c.}
\end{equation}
When the laser is far from resonance with the atom, the atom-laser interaction can be treated using second order perturbation theory within each $\{(J, m), \; |m| \leq J\}$ manifold.
Provided that the Zeeman splitting between the magnetic sublevels is much larger than the coupling induced by the atom-laser interaction, the electronic eigenstates remain unchanged at the lowest order, and the effect of the laser reduces to a shift of their energies proportional to the squared electric field amplitude:
\begin{equation}
    \label{eq:Stark_shift}
    V_{J,m}(\omega, \mathbf{u}) = - \alpha_{J,m}(\omega, \mathbf{u}) \frac{|\mathcal{E}|^2}{4} \; ,
\end{equation}
This energy shift is called the (dynamic) Stark shift, and the coefficient of proportionality $\alpha$ is called the (dynamic) polarizability.

Following a common usage, we decompose the polarizability into scalar ($\alpha_J^\text{s}$), vector ($\alpha_J^\text{v}$) and tensor ($\alpha_J^\text{t}$) polarizabilities:
\begin{multline}
    \alpha_{J,m}(\omega, \mathbf{u})
    = \alpha_J^\text{s}(\omega) - i\alpha_J^\text{v}(\omega) \frac{(\mathbf{u^\ast\!\times u}) \cdot \mathbf{J}}{2J} \\
    + \alpha_J^\text{t}(\omega) \frac{3 (\mathbf{u^\ast\!\cdot J})(\mathbf{u \cdot J}) - \mathbf{J}^2}{J(J - 1)} \; .
\end{multline}
The vector polarizability only contributes if the ellipticity of the polarization is non-zero, because otherwise $\mathbf{u}$ is a real vector and the cross product $\mathbf{u^\ast \times u}$ is zero.
In order to compute explicitly the scalar and vector products, we introduce the Cartesian basis $(\mathbf{e}_x, \mathbf{e}_y, \mathbf{e}_z)$, with the $z$-axis pointing along the magnetic field: $\mathbf{e}_z = \mathbf{B} / \|\mathbf{B}\|$.
One then obtains the following, more practical expression:
\begin{multline}
    \label{eq:polarizability}
    \alpha_{J,m}(\omega, \mathbf{u})
    = \alpha_J^\text{s}(\omega) + \alpha_J^\text{v}(\omega) \frac{m}{2J} C(\mathbf{u}) \\
    - \alpha_J^\text{t}(\omega) \frac{3\,m^2 - J(J+1)}{2J(2J-1)} D(\mathbf{u}) \; ,
\end{multline}
with
\begin{align}
    \label{eq:coeff_C}
    C(\mathbf{u}) &= 2 \mathfrak{Im}(u_x^\ast u_y^{\vphantom{\ast}}) \; , \\
    \label{eq:coeff_D}
    D(\mathbf{u}) &= 1 - 3 |u_z|^2 \; .
\end{align}
The coefficient $C$ quantifies the projection of the spin angular momentum of the light onto the quantization axis $z$: $C = \pm 1$ for $\sigma_\pm$ light, while $C = 0$ for $\pi$ light.
The scalar, vector and tensor polarizabilities are related to the irreducible components of the polarizability tensor by the following equalities \cite{LeKien2013}:
\begin{align}
    \label{eq:alpha_s}
    \alpha_J^\text{s} &= \frac{1}{\sqrt{3(2J + 1)}} \alpha_J^{(0)} \; , \\
    \label{eq:alpha_v}
    \alpha_J^\text{v} &= -\frac{\sqrt{2J}}{\sqrt{(J+1)(2J+1)}} \alpha_J^{(1)} \; , \\
    \label{eq:alpha_t}
    \alpha_J^\text{t} &= -\frac{\sqrt{2J(2J-1)}}{\sqrt{3(J+1)(2J+1)(2J+3)}} \alpha_J^{(2)} \; .
\end{align}
The irreducible components of the polarizability tensor can be calculated using the expression
\begin{widetext}
\begin{equation}
    \label{eq:alpha_tensor}
    \alpha_J^{(K)}(\omega) = (-1)^{K+J+1} \sqrt{2K+1} \sum_{J'} \bigg[ (-1)^{J'} \medmath{\begin{Bmatrix} 1 & K & 1 \\ J & J' & J \end{Bmatrix}} \, \frac{\left|\langle J'\| \mathbf{d}\|J \rangle\right|^2}{\hbar} \mathfrak{Re} \bigg( \frac{1}{\omega_{J'J} - \omega - i\gamma_{J'J} / 2} + \frac{(-1)^K}{\omega_{J'J} + \omega + i\gamma_{J'J} / 2} \bigg) \bigg]\; .
\end{equation}
\end{widetext}
In this equation, $\omega_{J'J} = \omega_{J'} - \omega_J$ is the difference between the energies of the states $J'$ and $J$ (divided by $\hbar$), $\gamma_{J'J} = \gamma_{J'} + \gamma_J$ is the sum of the inverse radiative lifetimes of the states $J'$ and $J$, and $\langle J'\|\mathbf{d}\|J \rangle$ is the reduced dipole matrix element and of the transition $J \leftrightarrow J'$.
The sum runs over all possible transitions from the state $J$ to higher or lower lying states $J'$, and we have used the notation $\begin{Bsmallmatrix} . & . & . \\ . & . & . \end{Bsmallmatrix}$ for Wigner's 6-$j$ symbol.
The contribution of the Zeeman shift to the state energies is completely negligible for the large laser detuning $\Delta = \omega - \omega_{J'J}$ we are interested in.
The reduced dipole matrix elements can be calculated from the transition rates 
\begin{equation}
    A_{J'J} = \frac{1}{(2J'+1)} \frac{\omega_{J'J}^3}{3 \pi \varepsilon_0 c^3\hbar} \left| \langle J' \| \mathbf{d} \| J \rangle \right|^2 \quad (\text{for } \omega_{J'}>\omega_J) \; ,
\end{equation}
and the inverse radiative lifetimes are equal to the sum of the transition rates towards lower lying states:
\begin{equation}
    \gamma_{J'} = \sum_{J} A_{J'J}  \quad (\text{for } \omega_{J'}>\omega_J) \; .
\end{equation}

In this work, we have measured the scalar, vector and tensor components of the differential polarizability between the ground state \gfull (label `g') and different sublevels $m$ of the excited state \efull (label `e') at \qty{1064.7}{\nm}:
\begin{multline}
    \label{eq:diff_pol}
    \Delta\alpha_{m}(\omega, \mathbf{u}) = \Delta\alpha^\text{s}(\omega) + \alpha^\text{v}_\text{e}(\omega) \frac{m}{2} C(\mathbf{u}) \\
    - \alpha^\text{t}_\text{e}(\omega) \frac{3\,m^2 - 2}{2} D(\mathbf{u}) \; ,
\end{multline}
with
\begin{equation}
    \Delta\alpha^\text{s} = \alpha^\text{s}_\text{e} - \alpha^\text{s}_\text{g} \; .
\end{equation}
Throughout the article, we will give the values of polarizabilities in units of $e^2 a_0^4 m_e/\hbar^2$ (atomic units, abbreviated \si{\au}), with $e$ the electron charge, $a_0$ the Bohr radius, and $m_e$ the electron mass.
We have also computed reference values using \crefrange{eq:alpha_s}{eq:diff_pol} and the most recent spectroscopic data available in the literature, which are listed in \cref{tab:1S0,tab:3P1}.
For the sake of completeness, we have included the ionic core polarizabilities (corrected by the core-valence contributions), $\alpha_g^\text{core} = \qty{5.3}{\au}$ and $\alpha_e^\text{core} = \qty{5.6}{\au}$ \cite{Mitroy2010,Cooper2018}, as frequency-independent contributions to the scalar polarizabilities
\footnote{The core excitations lie at wavelengths much shorter than the one typically used for optical trapping, which is why we consider the core polarizability as frequency-independent.}.
Our theoretical prediction at \qty{1064.7}{\nm} is: 
\begin{equation*}
    \Delta\alpha^\text{s} = \qty{-67}{\au} \; , \;\;
    \alpha^\text{v}_\text{e} = \qty{-172}{\au} \; , \;\; \text{and} \;\;
    \alpha^\text{t}_\text{e} = \qty{+17}{\au} \; .
\end{equation*}
According to these values, it should be possible to cancel the differential polarizability between the ground state and the excited sublevels $m = \pm 1$ for specific orientations of the polarization vector $\mathbf{u}$ with respect to the quantization axis.
One such configuration corresponds to a laser beam propagating in the same direction as the magnetic field ($+z$), and a polarization such that
\begin{equation*}
    C_\text{magic} = \pm \frac{\alpha_\text{e}^\text{t} - 2 \Delta\alpha_\text{s}}{\alpha_\text{e}^\text{v}} = \mp \num{0.88} \;\; \text{for } m = \pm 1 \; .
\end{equation*}
As already mentioned, the existence of a magic polarization was also predicted recently in \cite[pp.~39-40]{MadjarovThesis} using a slightly different approach: there, only the strongest transitions were explicitly included in \cref{eq:alpha_tensor}, and offsets were added to the scalar polarizabilities to take into account all other transitions, with their values adjusted to reproduce the magic configurations at \qtylist{813;914}{\nm}.
In spite of the convergence of these predictions, it is extremely difficult to assess their accuracy because the spectroscopic data they rely on have very different origins, and one can hardly aggregate their uncertainties (which is why we decided not to provide an uncertainty on the predicted values).
A direct experimental measurement of the differential polarizability therefore remains essential.

\section{Experimental protocol%
\label{sec:protocol}}

\begin{figure}
    \centering
    \includegraphics{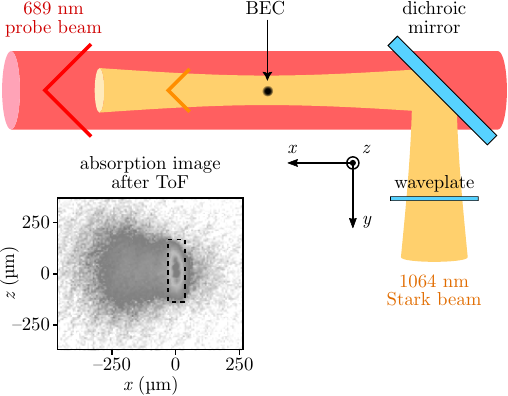}
    \caption{A top view representation of the experimental setup.
    The atomic sample was a Bose--Einstein condensate (BEC) of \nuclide[84]{Sr} atoms.
    A laser beam with a wavelength of \qty{1064.7}{\nm}, a $1/e^2$-radius of \qty{285}{\um}, and a power close to \qty{20}{\W} induced a Stark shift on the electronic levels.
    We call it the `Stark' beam.
    We varied its polarization using a half-wave or a quarter-wave plate.  
    While the Stark beam was on, we excited the intercombination transition at \qty{689}{\nm} using a $\sim$~\qty{1}{\cm} large beam, which we call the `probe' beam.
    The orientation of the atomic dipoles is imposed by a \qty{1.5}{\Gauss} bias magnetic aligned either along $x$ or along $y$ (not shown). 
    After a \qty{18}{\ms} long time of flight (ToF), we recorded an absorption image of the atomic distribution.
    When the probe beam was on resonance, a fraction of the atoms was kicked away from the BEC, which we could directly measure by integrating the signal in a region of interest encompassing the BEC mode (dashed rectangle in the camera image).
    We draw the reader's attention on the fact that the Cartesian coordinate system used to describe the experiment is fixed by the geometry of the setup, not by the orientation of the magnetic field. This is in contrast with the convention used in \cref{sec:theory}.
    }
    \label{fig:diagram}
\end{figure}

Before going into the details of our experimental protocol, we first briefly explain our strategy, which is also schematically depicted in \cref{fig:diagram}.
We shined a powerful \qty{1064}{\nm} laser beam onto a Bose--Einstein condensate (BEC) of \nuclide[84]{Sr} atoms to induce a Stark shift, and probed the intercombination transition with another \qty{689}{\nm} laser beam.
After a time of flight, we recorded an absorption image and measured the fraction of atoms which were expelled from the BEC mode by the probe beam.
By repeating the measurement for different frequencies of the probe beam, we determined the position of the shifted resonance.
Then, we obtained the differential Stark shift by subtracting the position found in the absence of Stark beam, and deduced the differential polarizability after dividing by the calibrated Stark beam intensity.
In order to separate the scalar, vector, and tensor components of the polarizability, we repeated this procedure for different orientations of the atomic dipole (imposed by an external magnetic field), and different polarizations of the Stark beam.

Here, we describe each step more specifically.
The BEC contained \numrange{1e5}{4e5} \nuclide[84]{Sr} atoms and was prepared in one anti-node of a vertical ($z$-axis) optical lattice formed by the intersection of two \qty{1064}{\nm} laser beams with a relative angle of \qty{12}{\degree}.
The trap confinement frequencies at the end of the evaporation were \qty{1.8}{\kHz} in the vertical direction and \qty{12}{\Hz} in both horizontal directions.
The optical density of this sample was so high that probing the intercombination transition at this stage resulted in strongly asymmetric line shapes, unsuitable for the precision measurements we were aiming for.
In order to reduce the optical density, we proceeded in two steps. 
Firstly, we suddenly increased the intensity of the trapping laser beams by a factor 4 and held the gas for a quarter of the new vertical oscillation period in order to convert the initial potential energy into kinetic energy in the vertical direction. The horizontal dynamics was negligible during this compression step.
Then, we switched off the optical trap and let the gas expand freely during \qty{1}{\ms}, after which the vertical dimension had increased to \qty{12}{\um}, while the horizontal dimensions remained approximately constant at \qty{50}{\um}, both being defined as the full width at half maximum (FWHM).
The peak atomic density was then of the order \qty{5e12}{\cm^{-3}}, corresponding to an optical density of the order of \num{30} on resonance.

At this stage, we shined the Stark beam and simultaneously excited the \gshort--\eshort intercombination transition with the probe beam for a duration of \qty{800}{\us}.
The Stark beam was aligned on the position of the atoms after the first \qty{1}{\ms} expansion, and propagated in the horizontal plane along the $x$-axis, see \cref{fig:diagram}.
The beam had a $1/e^2$-radius of \qty{285}{\um} and a power close to \qty{20}{\W}.
With such parameters, the confining effect of the Stark beam was too weak to hold the atoms against gravity, and the atoms kept falling while they were being probed. However, the displacement of the sample during the \qty{800}{\us} probe duration was about \qty{10}{\um}, which is negligible.
We have carefully calibrated the Stark beam intensity at the position of the atoms to be $I = \varepsilon_0 c |\mathcal{E}|^2 / 2 = \qty{147 \pm 5}{\W/\mm^2}$ ($\varepsilon_0$ is the vacuum dielectric permittivity and $c$ the speed of light).
The uncertainty should be understood as peak-to-peak, and it aggregates the uncertainties on the relative position between the atoms and the beam center, on the beam profile, and on the beam power.
More details are given in appendix \ref{app:calib-intensity}.
The polarization of the Stark beam, initially linear, was tuned either with a quarter-wave plate or with a half-wave plate, depending on the polarizability component which we wanted to determine.
We have verified experimentally that the birefringence of the optical elements located in-between the waveplate and the atoms had a negligible effect on the polarization of the Stark beam (below \qty{1}{\percent}).

The probe beam was parallel to the Stark beam, and it had a diameter of the order of \qty{1}{\cm}, much larger than the size of the atomic distribution, see \cref{fig:diagram}.
Its intensity was set approximately equal to the saturation intensity of the intercombination transition (\qty{3}{\uW/\cm^2}).

We used three pairs of coils surrounding the vacuum chamber to compensate for the earth magnetic field and apply a bias field of $B = \qty{1.5}{\Gauss}$ throughout the probing phase.
The Zeeman splitting $g_J \mu_\text{B} B / h \simeq \qty{3}{\MHz}$ is more than 20 times larger than the Stark shift ($g_J$ is the Landé factor, $\mu_\text{B}$ the Bohr magneton, and $h$ the Planck constant), consequently, the orientation of the atomic dipoles was effectively imposed by the orientation of the magnetic field, and we could easily resolve the transitions to the different $\eshort, m$ states.
The magnetic field was oriented along $x$ (i.e., parallel to the Stark beam) when using the quarter-wave plate, and along $y$ (i.e., perpendicular to the Stark beam) when using the half-wave plate.
The exact procedure used for aligning the magnetic field is described in appendix \ref{app:alignement-mag-field}.

After the probing phase, we let the gas freely fall and expand for another \qty{18}{\ms} in order to spatially separate the atoms which have absorbed a photon from the probe beam from those which have not.
Then, we recorded an absorption image using the broad \gshort--$\tensor[^1]{\mathrm{P}}{_1}$ transition at \qty{461}{\nm}, with the line of sight along the $y$ axis, see \cref{fig:diagram}.
To locate the atomic resonance, we measured the fraction of atoms remaining in a region of interest centered on the BEC mode as we scanned the probe beam frequency, see \cref{fig:diagram} and the inset in \cref{fig:hwp_dls}.
Due to the high optical density, the intercombination line was broadened to around \qty{60}{\kHz} FWHM.
By repeating the same experiment in the absence of the Stark beam and subtracting the position of the free-space resonance from that of the shifted resonance, we directly obtained the differential Stark shift,
\begin{equation}
    \Delta V_m(\mathbf{u}) = V_{\text{e},m}(\mathbf{u}) - V_{\text{g}}(\mathbf{u}) = - \Delta\alpha_m(\mathbf u) \frac{I}{2\varepsilon_0 c} \; .
\end{equation}

\section{Results%
\label{sec:results}}

\begin{figure}
    \centering
    \includegraphics{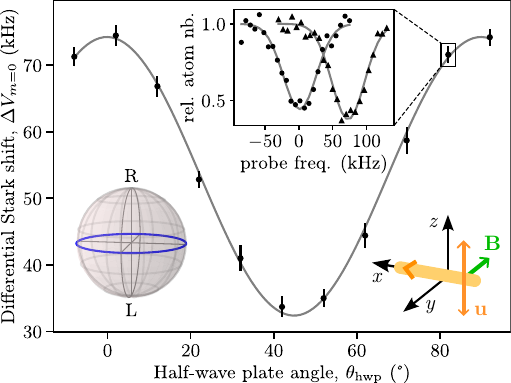}
    \caption{Differential Stark shift of the $\gshort \text{--} \eshort, m=0$ transition as we rotate the direction of the linear Stark beam polarization.
    The direction of the atomic dipole ($y$) is fixed by the external magnetic field $\mathbf{B}$ and is perpendicular to the Stark beam propagation axis ($x$).
    The direction of the linear Stark beam polarization $\mathbf{u}$ is rotated using a half-wave plate, whose orientation is measured by the angle $\theta_\text{hwp}$.
    The dots with error bars represent the measured differential Stark shifts.
    The blue line on the Poincaré sphere represents the trajectory of the Stark beam polarization as we vary $\theta_\text{hwp}$.
    For each orientation of the waveplate, we measure the relative atom number in the BEC mode as we scan the probe beam frequency relative to the expected free-space resonance, both in the presence and in the absence of the Stark beam (triangles and dots in the inset, respectively).
    Then, we fit the spectra with an inverted Gaussian model to determine the resonances positions (gray lines in the inset), and take the difference of the two values to obtain the differential Stark shift.
    The uncertainties on the measured differential Stark shift (vertical error bars on the main graph) reflect the fit uncertainties on the resonance positions.
    The uncertainty on the orientation of the half-wave plate is estimated to be \qty{\pm 0.5}{\degree} and is not visible on the scale of the graph.
    The sinusoidal gray line in the main graph is the result of a fit of the measured differential Stark shift by a sinusoidal model using orthogonal distance regression.
    }
    \label{fig:hwp_dls}
\end{figure}

We have performed two complementary sets of measurements.
In the first set, we probed the transition to $\eshort, m=0$ with a linear Stark beam polarization, and fixed the orientation of the atomic dipole along $y$, i.e., perpendicular to the Stark beam propagation axis, using the external magnetic field.
The direction of the Stark beam polarization was then varied by rotating a half-wave plate.
In this configuration, the excited state has no vector polarizability, and its tensor polarizability oscillates sinusoidally with the angle of the waveplate, such that the differential polarizability reduces to
\begin{equation}
    \label{hwp_model}
    \Delta\alpha_{m=0}(\theta_\text{hwp}) = \Delta\alpha^\text{s} - \frac{1}{2} \alpha^\text{t}_\text{e} - \frac{3}{2} \cos\left[ 4(\theta_\text{hwp} - \theta_{0,\text{hwp}}) \right] \,\alpha^\text{t}_\text{e} \; ,
\end{equation}
where $\theta_\text{hwp}$ is the angle between one optic axis of the waveplate and a reference orientation, and $\theta_{0,\text{hwp}}$ is the angle between the reference orientation and the initial linear polarization of the Stark beam.
The differential Stark shifts measured for different orientations of the half-wave plate are displayed as dots with error bars in \cref{fig:hwp_dls}, and match very well with the expected sinusoidal profile.
The trajectory followed the Stark beam polarization as we rotated the waveplate is represented by the blue line is located along the equator of the Poincaré sphere, corresponding to the linear polarizations in the transverse plane $yz$
\footnote{We adopt the point of view of the source to define the right- and left-handed circularly polarized light, respectively denoted R and L. With this convention, when the light beam propagation axis is aligned with the quantization axis, right-handed circularly polarized light is $\sigma^+$ and left-handed circularly polarized light is $\sigma^-$.}.
The graph in inset shows the raw spectra from which we extract the differential Stark shift for $\theta_\text{hwp} = \qty{82}{\degree}$: the triangles correspond to the shifted resonance, the dots to the free-space resonance, and the gray lines are the fitted inverted Gaussian profiles which we used to determine the resonance positions.
The vertical error bars in \cref{fig:hwp_dls} is the root sum of squares of the standard deviations on the fitted resonance positions.
We have estimated the uncertainty on the angle $\theta_\text{hwp}$ to \qty{\pm 0.5}{\degree}, which is too small to be visible on the scale of the graph.
Using orthogonal distance regression (gray line in the main graph), we find the amplitude and baseline of the oscillation to be, respectively, \qty{20.9 \pm 0.3}{\kHz} and \qty{53.3 \pm 0.2}{\kHz}, where the uncertainties correspond to one standard deviation.
Dividing by the independently calibrated Stark beam intensity, we obtain
\begin{align*}
    \alpha^\text{t}_\text{e} &= \qty{+20.2 \pm 0.9}{\au} \; ,\\
    \Delta\alpha^\text{s} - \frac{1}{2}\alpha^\text{t}_\text{e} &= \qty{-77.4 \pm 2.7}{\au} \; .
\end{align*}
The uncertainties on these values combine the statistical uncertainty on the corresponding regression parameter, and the systematic uncertainty on the Stark beam intensity.
We give the two uncertainties separately in \cref{tab:results}.

\begin{figure}
    \centering
    \includegraphics{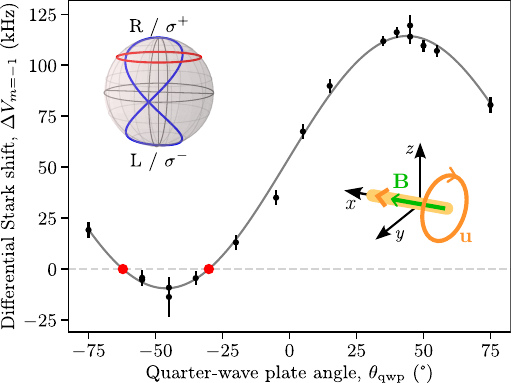}
    \caption{Differential Stark shift of the \gshort--$\eshort, m=-1$ transition as we vary the ellipticity of the Stark beam polarization.
    The direction of the atomic dipole ($x$) is fixed by the external magnetic field $\mathbf{B}$ and is parallel to the Stark beam propagation axis.
    The ellipticity of the Stark beam polarization $\mathbf{u}$ is controlled by the orientation of a quarter-wave plate (angle $\theta_\text{qwp}$).
    The dots, the error bars and the line in the main graph have the same meaning as in \cref{fig:hwp_dls}.
    The uncertainty on the orientation of the half-wave plate is estimated to be \qty{\pm 1}{\degree}.
    The differential Stark shift is found to be zero for a magic ellipticity $C_\text{magic} = \num{+0.847 \pm 0.023}$ (red dots), corresponding to the latitude \qty{+57.9}{\degree} on the Poincaré sphere (red circle).
    }
    \label{fig:qwp_dls}
\end{figure}

In the second set of measurements, we probed the transition to $\eshort, m=-1$ with an elliptical Stark beam polarization, and fixed the orientation of the atomic dipole along $x$, i.e., parallel to the Stark beam propagation axis, using the external magnetic field.
We varied the ellipticity of the Stark beam polarization by rotating a quarter-wave plate.
In this configuration, the tensor polarizability of the excited state is constant, and the vector polarizability oscillates with the angle of the waveplate, leading to
\begin{equation}
    \label{eq:qwp_model}
    \Delta\alpha_{m=-1}(\theta_\text{qwp}) = \Delta\alpha^\text{s} - \frac{1}{2}\alpha^\text{t}_\text{e} + \frac{1}{2}\sin[2(\theta_\text{qwp} - \theta_{0,\text{qwp}})] \, \alpha^\text{v}_\text{e} \; .
\end{equation}
Our measurements of the differential Stark shift for different orientations of the quarter-wave plate are again in excellent agreement with this model, see \cref{fig:qwp_dls}.
Note that the uncertainty on the waveplate angle was \qty{\pm 1}{\degree} in this configuration.
The trajectory of the Stark beam polarization on the Poincaré sphere goes from linear (equatorial plane) when one of the principal axes of the waveplate is aligned with the initial polarization, to right-handed circular (equivalently $\sigma^+$, north pole) or left-handed circular (equivalently $\sigma^-$, south pole), when the optic axes are at \qty{45}{\degree} from the initial polarization.
The fitted values of the amplitude and baseline of the oscillation are, respectively, \qty{61.9 \pm 0.9}{\kHz} and \qty{52.5 \pm 0.8}{\kHz}.
Dividing by the Stark beam intensity yields
\begin{align*}
    \alpha^\text{v}_\text{e} &= \qty{-180 \pm 7}{\au} \; , \\
    \Delta\alpha^\text{s} - \frac{1}{2}\alpha^\text{t}_\text{e} &= \qty{-76.2 \pm 2.9}{\au} \; . \\
\end{align*}
Note that the static component $\Delta\alpha^\text{s} - \frac{1}{2}\alpha^\text{t}_\text{e}$ measured in the first and second configurations match to within \qty{2}{\percent}, although the two sets of measurements have been performed at a one-week interval, and the Stark beam alignment was performed independently for each set.

Remarkably, the oscillation of the differential Stark shift intercepts zero for two positions of the quarter-wave plate (horizontal dashed line, red dots), corresponding to a `magic' polarization characterized by
\begin{equation*}
    C_\text{magic} = \frac{\alpha_\text{e}^\text{t} - 2 \Delta\alpha_\text{s}}{\alpha_\text{e}^\text{v}} = \num{+0.847 \pm 0.023} \; ,
\end{equation*}
and to the latitude \qty{+57.9}{\degree} (red circle) on the Poincaré sphere.
We stress that the measurement of $C_\text{magic}$ is free of the systematic uncertainty related to the Stark beam intensity.

\begin{table}
    \centering
    \begin{ruledtabular}
        \begin{tabular}{c S[table-format=+3.1] S[table-format=1.1] S[table-format=2.1] c c}
        & {measured} &  \multicolumn{2}{c}{std. dev.} & {predicted} & unit \\
        \cmidrule(lr){3-4}
        & & {stat.} & {syst.} & & \\
        \midrule
        $\alpha_\text{e}^\text{v}$ & -180 & 3 & 6 & \tablenum[table-format=+3]{-172} & \multirow{4}{*}{\unit{\au}} \\[2pt]
        $\alpha^\text{t}_\text{e}$ & +20.2 & 0.6 & 0.7 & \tablenum[table-format=+3]{+17} & \\[2pt]
        \multirow{2}{*}{$\Delta\alpha^\text{s} - \frac{1}{2} \alpha^\text{t}_\text{e}$} & -77.4 & 0.7 & 2.6 & \multirow{2}{*}{\tablenum[table-format=+3]{-76}} & \\
        & -76.2 & 1.3 & 2.6 & & \\
        \midrule
        $|C_\text{magic}|$ & {0.847} & {0.023} & & {0.88} & \\
        \end{tabular}
    \end{ruledtabular}
    \caption{Summary of our measurements of the components of the differential polarizability of strontium at \qty{1064.7}{\nm}.
    The polarizability values are given in units of $e^2 a_0^4 m_e/\hbar^2$ (atomic units).
    The standard deviations for these values are decomposed into a statistical uncertainty linked to the differential Stark shift measurements, and a systematic uncertainty associated with the calibration of the Stark beam intensity.
    The predicted polarizability values were obtained using \cref{eq:alpha_s,eq:alpha_v,eq:alpha_t,eq:alpha_tensor} and the spectroscopic data from \cref{tab:1S0,tab:3P1}.
    }
    \label{tab:results}
\end{table}

\section{Conclusion}

We have measured the scalar, vector and tensor components of the differential dynamic polarizability of the \gshort--\eshort intercombination transition of strontium at a wavelength of \qty{1064.7}{\nm}.
We have also identified a magic ellipticity \num{\pm 0.85 \pm 0.02} of the polarization at which the differential polarizability on the $m = \pm 1$ states vanishes.
The existence of this magic configuration has important implications for laser cooling optically trapped atoms because it gives access to the three main types of cooling mechanisms (sideband cooling, attractive and repulsive Sisyphus cooling), at a wavelength were powerful laser sources exist.
In particular, tuning to the sideband or the attractive Sisyphus cooling limits would relax the constraint on the trap depth which is imposed by the repulsive Sisyphus cooling.
The laser power could then be used instead to increase the number of optical traps, or their volume.
Our observations also provide a new benchmark for atomic models of strontium, which should also be beneficial to precision experiments.

\begin{acknowledgments}

The authors express their gratitude to Ana\"is Molineri, Clémence Briosne-Fr\'ejaville, Sayali Shevate and Florence Nogrette for their contribution to the design and construction of the experimental apparatus, and to Isabelle Bouchoule for steady and fruitful discussions.
The authors also thank Maxence Lepers and Mona Ghazal for their critical reading of the manuscript.
This work has been supported by Region Île-de-France in the framework of DIM QuanTiP.
This project has received funding from the European Research Council (ERC) under the European Union’s Horizon 2020 research and innovation programme (grant agreement No 679408).

\end{acknowledgments}

\appendix

\section{Spectroscopic data for the computation of the polarizabilities}

The lists of all transitions included in our computation of the polarizability of the states \gfull and \efull are given in \cref{tab:1S0,tab:3P1}.
{\begin{table}
   \centering
   \begin{ruledtabular}
      \begin{tabular}{cS[table-format=3.2,round-mode=places,round-precision=2]S[table-format=1.3,round-mode=places,round-precision=3]c}
         $J'$  &  {wavelength}  &  {$\left| \langle J' \| \mathbf{d} \| J \rangle \right|$} &  source   \\
         & {(\unit{\nm})} & {($e a_0$)} & \\
         \midrule
         \ensuremath{5s5p\,\tensor[^3]{\mathrm{P}}{_1}}  &  689.46   &  0.151000 &  \cite{NIST-ASD,Cooper2018} \\
         %\ensuremath{5s5p\,\tensor[^1]{\mathrm{P}}{_1}} &  460.87   & 5.394400  &  \cite{NIST-ASD}            \\ % Ancienne valeur
         \ensuremath{5s5p\,\tensor[^1]{\mathrm{P}}{_1}}  &  460.87   & 5.248     &  \cite{Cooper2018}          \\
         \ensuremath{5s6p\,\tensor[^1]{\mathrm{P}}{_1}}  &  293.27   & 0.266460  &  \cite{NIST-ASD}            \\
         \ensuremath{5s6p\,\tensor[^3]{\mathrm{P}}{_1}}  &  295.26   & 0.108000  &  \cite{Ruczkowski2016}      \\
         \ensuremath{4d5p\,\tensor[^3]{\mathrm{P}}{_1}}  &  268.08   & 0.043000  &  \cite{Ruczkowski2016}      \\
         \ensuremath{5s7p\,\tensor[^1]{\mathrm{P}}{_1}}  &  257.02   & 0.360560  &  \cite{NIST-ASD}            \\
         \ensuremath{5s7p\,\tensor[^3]{\mathrm{P}}{_1}}  &  253.64   & 0.053000  &  \cite{Ruczkowski2016}      \\
         \ensuremath{4d5p\,\tensor[^1]{\mathrm{P}}{_1}}  &  242.88   & 0.600000  &  \cite{NIST-ASD}            \\
         \ensuremath{5s8p\,\tensor[^3]{\mathrm{P}}{_1}}  &  239.69   & 0.033000  &  \cite{Ruczkowski2016}      \\
         \ensuremath{5s8p\,\tensor[^1]{\mathrm{P}}{_1}}  &  235.50   & 0.591610  &  \cite{NIST-ASD}            \\
         \ensuremath{5s9p\,\tensor[^3]{\mathrm{P}}{_1}}  &  232.61   & 0.017000  &  \cite{Ruczkowski2016}      \\
         \ensuremath{5s9p\,\tensor[^1]{\mathrm{P}}{_1}}  &  230.80   & 0.457170  &  \cite{NIST-ASD}            \\
         \ensuremath{5s10p\,\tensor[^3]{\mathrm{P}}{_1}} &  228.51   & 0.012000  &  \cite{Ruczkowski2016}      \\
         \ensuremath{5s10p\,\tensor[^1]{\mathrm{P}}{_1}} &  227.59   & 0.346410  &  \cite{NIST-ASD}            \\
         \ensuremath{5s11p\,\tensor[^3]{\mathrm{P}}{_1}} &  225.91   & 0.009000  &  \cite{Ruczkowski2016}      \\
         \ensuremath{5s11p\,\tensor[^1]{\mathrm{P}}{_1}} &  225.40   & 0.251000  &  \cite{NIST-ASD}            \\
         \ensuremath{5s12p\,\tensor[^1]{\mathrm{P}}{_1}} &  223.83   & 0.200000  &  \cite{NIST-ASD}            \\
         \ensuremath{5s13p\,\tensor[^1]{\mathrm{P}}{_1}} &  222.70   & 0.160310  &  \cite{NIST-ASD}            \\
         \ensuremath{5s14p\,\tensor[^1]{\mathrm{P}}{_1}} &  221.85   & 0.137480  &  \cite{NIST-ASD}            \\
         \ensuremath{5s15p\,\tensor[^1]{\mathrm{P}}{_1}} &  221.20   & 0.118320  &  \cite{NIST-ASD}            \\
         \ensuremath{5s16p\,\tensor[^1]{\mathrm{P}}{_1}} &  220.69   & 0.100000  &  \cite{NIST-ASD}            \\
         \ensuremath{5s17p\,\tensor[^1]{\mathrm{P}}{_1}} &  220.29   & 0.088882  &  \cite{NIST-ASD}            \\
         \ensuremath{5s18p\,\tensor[^1]{\mathrm{P}}{_1}} &  219.96   & 0.081240  &  \cite{NIST-ASD}            \\
         \ensuremath{5s19p\,\tensor[^1]{\mathrm{P}}{_1}} &  219.69   & 0.070711  &  \cite{NIST-ASD}            \\
         \ensuremath{5s20p\,\tensor[^1]{\mathrm{P}}{_1}} &  219.47   & 0.063325  &  \cite{NIST-ASD}            \\
      \end{tabular}
   \end{ruledtabular}
   \caption{Spectroscopic data used to compute the polarizability of the state $J \equiv \gfull$.}
   \label{tab:1S0}
\end{table}}
{\begin{table}
   \centering
   \begin{ruledtabular}
      \begin{tabular}{cS[table-format=3.2,round-mode=places,round-precision=2]S[table-format=1.3,round-mode=places,round-precision=3]c} 
         $J'$  &  {wavelength}  &  {$\left| \langle J' \| \mathbf{d} \| J \rangle \right|$}  &  source   \\
         & {(\unit{\nm})} & {($e a_0$)} & \\
         \midrule
         \ensuremath{5s4d\,\tensor[^3]{\mathrm{D}}{_1}}  &  2735.98  &   2.32200 &  \cite{Cooper2018}          \\
         \ensuremath{5s4d\,\tensor[^3]{\mathrm{D}}{_2}}  &  2692.51  &   4.01900 &  \cite{Cooper2018}          \\
         \ensuremath{5s4d\,\tensor[^1]{\mathrm{D}}{_2}}  &  1771.48  &   0.19000 &  \cite{Cooper2018}          \\
         \ensuremath{5s^2\,\tensor[^1]{\mathrm{S}}{_0}}  &  689.46   &   0.15100 &  \cite{NIST-ASD,Cooper2018} \\
         \ensuremath{5s6s\,\tensor[^3]{\mathrm{S}}{_1}}  &  688.04   &   3.42500 &  \cite{Cooper2018}          \\
         \ensuremath{5s6s\,\tensor[^1]{\mathrm{S}}{_0}}  &  621.62   &   0.04500 &  \cite{Cooper2018}          \\
         \ensuremath{5s5d\,\tensor[^1]{\mathrm{D}}{_2}}  &  494.49   &   0.06100 &  \cite{Cooper2018}          \\
         \ensuremath{5s5d\,\tensor[^3]{\mathrm{D}}{_1}}  &  487.73   &   2.00900 &  \cite{Cooper2018}          \\
         \ensuremath{5s5d\,\tensor[^3]{\mathrm{D}}{_2}}  &  487.38   &   3.67300 &  \cite{Cooper2018}          \\
         \ensuremath{5p^2\,\tensor[^3]{\mathrm{P}}{_0}}  &  483.35   &   2.65700 &  \cite{Cooper2018}          \\
         \ensuremath{5p^2\,\tensor[^3]{\mathrm{P}}{_1}}  &  478.56   &   2.36200 &  \cite{Cooper2018}          \\
         \ensuremath{5p^2\,\tensor[^3]{\mathrm{P}}{_2}}  &  472.37   &   2.86500 &  \cite{Cooper2018}          \\
         \ensuremath{5p^2\,\tensor[^1]{\mathrm{D}}{_2}}  &  445.30   &   0.22800 &  \cite{Cooper2018}          \\
         \ensuremath{5p^2\,\tensor[^1]{\mathrm{S}}{_0}}  &  441.38   &   0.29100 &  \cite{Cooper2018}          \\
         \ensuremath{5s7s\,\tensor[^3]{\mathrm{S}}{_1}}  &  436.30   &   0.92100 &  \cite{Cooper2018}          \\
         \ensuremath{5s7s\,\tensor[^1]{\mathrm{S}}{_0}}  &  417.71   &   0.25000 &  \cite{Ruczkowski2016}      \\
         \ensuremath{5s6d\,\tensor[^3]{\mathrm{D}}{_1}}  &  397.12   &   0.98635 &  \cite{Zhou2010}            \\
         \ensuremath{5s6d\,\tensor[^3]{\mathrm{D}}{_2}}  &  397.04   &   1.70840 &  \cite{Zhou2010}            \\
         \ensuremath{5s8s\,\tensor[^3]{\mathrm{S}}{_1}}  &  380.85   &   0.47880 &  \cite{Zhou2010}            \\
         \ensuremath{5s7d\,\tensor[^3]{\mathrm{D}}{_1}}  &  365.50   &   0.66100 &  \cite{Zhou2010}            \\
         \ensuremath{5s7d\,\tensor[^3]{\mathrm{D}}{_2}}  &  365.43   &   1.14480 &  \cite{Zhou2010}            \\
         \ensuremath{5s9s\,\tensor[^3]{\mathrm{S}}{_1}}  &  357.82   &   0.32350 &  \cite{Zhou2010}            \\
         \ensuremath{5s8d\,\tensor[^3]{\mathrm{D}}{_1}}  &  350.11   &   0.47950 &  \cite{Zhou2010}            \\
         \ensuremath{5s8d\,\tensor[^3]{\mathrm{D}}{_2}}  &  350.07   &   0.83060 &  \cite{Zhou2010}            \\
         \ensuremath{5s10s\,\tensor[^3]{\mathrm{S}}{_1}} &  345.74   &   0.24000 &  \cite{Zhou2010}            \\
         \ensuremath{5s9d\,\tensor[^3]{\mathrm{D}}{_2}}  &  341.26   &   0.64300 &  \cite{Zhou2010}            \\
         \ensuremath{5s9d\,\tensor[^3]{\mathrm{D}}{_1}}  &  341.26   &   0.37120 &  \cite{Zhou2010}            \\
         \ensuremath{4d^2\,\tensor[^3]{\mathrm{P}}{_0}}  &  333.09   &   1.68000 &  \cite{Ruczkowski2016}      \\
         \ensuremath{4d^2\,\tensor[^3]{\mathrm{P}}{_1}}  &  332.31   &   1.72000 &  \cite{Ruczkowski2016}      \\
         \ensuremath{4d^2\,\tensor[^3]{\mathrm{P}}{_2}}  &  330.84   &   2.21000 &  \cite{Ruczkowski2016}      \\
      \end{tabular}
   \end{ruledtabular}
   \caption{Spectroscopic data used to compute the polarizability of the state $J \equiv \efull$.}
   \label{tab:3P1}
\end{table}
}

\section{Calibration of the Stark beam intensity%
\label{app:calib-intensity}}

We took a great care in calibrating the value of the Stark beam intensity at the position of the atoms, because it is a major source of (systematic) uncertainty in our measurement of differential polarizability.
The first step was to measure the total power $P$ in the Stark beam using a power-meter.
To do so, we have calibrated with three different power meters the linear relationship between the absolute power and the voltage at the output of a photodiode which was monitoring a fraction of the Stark beam leaking from the rear side of a mirror.
Then, we extrapolated the power used in the experiments using the average of these calibrations to obtain the estimate $P = \qty{19.8}{\W}$, with an uncertainty of \qty{\pm 1.5}{\percent} (one standard deviation).

To compute the peak intensity of the beam, we recorded an image of the beam profile at the position of its waist, and we divided the count number of the brightest pixel by both the total count number and the surface of a pixel.
This procedure was necessary because the beam profile showed some deviations from a purely Gaussian mode, most noticeably in the tails.
In order to minimize the systematic errors, we reduced the background counts to a negligible amount, verified the linearity of the sensor at the level of the peak intensity, and measured the pixel size using a calibrated microscope.
We found a peak intensity of \qty{150}{\W/\mm^2} for $P = \qty{19.8}{\W}$.
Taking into account the uncertainty on the exact position of the atoms along the beam, the uncertainty on the pixel size, the dependence of the total count number on the choice of the integration area, and the uncertainty on the laser power, we estimate the uncertainty on the peak intensity to be \qty{\pm 2.5}{\percent}.
This number should be considered a conservative estimate as we could not make a statistical analysis for all sources of uncertainty.

Finally, we have also considered the effect of the relative position between the atoms and the point where the Stark beam reaches its peak intensity.
Here, we took two effects into account: the finite size of the atomic sample and the accuracy of the beam alignment.
After the \qty{1}{\ms} free expansion, the atomic sample reaches a full width at half maximum of \qty{50}{\um} in the $x$ and $y$ directions, and \qty{12}{\um} in the $z$ direction, while the Stark beam has a $1/e^2$ radius of \qty{285}{\um} in the $yz$ plane.
Modelling the atomic distribution by an axisymmetric Thomas--Fermi profile in the $xy$ plane, and by a Gaussian along $z$, we conclude that the average Stark beam intensity seen by the atoms is \qty{1}{\percent} lower than the peak intensity.

Our procedure to align the Stark beam onto the center of the atomic sample was to maximize the differential Stark shift for the state with $m=-1$ and a $\sigma^+$ polarization, that is, when the differential Stark shift was the largest.
We estimate that we could superimpose the center of the beam with the center of the sample to \qty{\pm 30}{\um}, taking into account the \qty{10}{\um} displacement of the atoms during the probe phase.
We therefore conclude that the average Stark beam intensity at the position of the atoms was between \num{0.95} and \num{0.99} times the peak intensity.
This range should again be considered a conservative estimate.

Putting together the estimate of the power of the Stark beam, of the peak intensity, and of the relative position between the atoms and the beam, we arrive at the following estimate for the Stark beam intensity: $I = \qty{147 \pm 5}{\W/\mm^2}$, where the uncertainty was rounded to the next upper integer.

\section{Relative orientation of the magnetic field and Stark beam propagation axis%
\label{app:alignement-mag-field}}

In our experiment, the orientation of the bias magnetic field (along $x$ or $y$) defines the quantization axis and serves as a reference to decompose the polarization vector.
It is therefore essential to precisely assess the relative orientation between the bias magnetic field and the Stark beam propagation axis.
For example, when measuring the differential Stark shift of the \gshort--$\eshort, m=-1$ as a function of the ellipticity of the Stark beam polarization (second configuration, \cref{eq:qwp_model} and \cref{fig:qwp_dls}), we assumed that the Stark beam was exactly parallel to the bias magnetic field.
If there was instead an angle $\beta$ between the Stark beam and the bias magnetic field, the amplitude of the oscillation would be reduced by a factor $\cos^2(\beta)$ with respect to the case $\beta = 0$, which would lead to an underestimation of the vector component $\alpha_e^\text{v}$.

The bias magnetic field was generated by three pairs of coils wound around the viewports of the vacuum chamber, with their axes defining the $x$, $y$ and $z$ axes.
The exact angle between the Stark beam and the coil axes has not been measured.
However, because the Stark beam was centered on the viewports of the $x$ axis, the angle between the Stark beam propagation axis and the magnetic field produced by the pair of coils on the $x$ axis must be within $\qty{\pm 1.5}{\degree}$ (conservative estimation).

The ambient magnetic field was compensated by scanning the probe beam frequency around the free-space frequency of the intercombination transition and tuning the currents in the coils until the lines corresponding to the three magnetic sublevels of the excited state merged.
The resulting line had a FWHM of \qty{50}{\kHz}, corresponding to a residual ambient magnetic lower than \qty{0.025}{\Gauss}.
Consequently, the orientation sum of \qty{1.5}{\Gauss} bias magnetic field and this residual ambient magnetic field should be within \qty{\pm 1}{\degree} of the coil's axis for all configurations.

In the end, the discrepancy between the relative orientation of the bias magnetic field and Stark beam propagation axis is therefore lower than \qty{3}{\degree}, and the resulting uncertainty on the different components of the differential polarizability is negligible in front of the other sources of uncertainty ($\cos(\qty{3}{\degree}) = \num{0.997}$).

\bibliography{../bibliography.bib}

\end{document}